# Non-Adiabatic Electronic and Vibrational Ring-Opening Dynamics resolved with Attosecond Core-Level Spectroscopy

S. Severino[1,†], K.M. Ziems[2,†], M. Reduzzi[1], A. Summers[1], H.-W. Sun[1], Y.-H. Chien[1], S. Gräfe[2,3,4], J. Biegert[1,5,*]

[1] ICFO - Institut de Ciencies Fotoniques, The Barcelona Institute of Science and Technology, 08860 Castelldefels (Barcelona), Spain
[2] Institute of Physical Chemistry and Max Planck School of Photonics, Friedrich-Schiller-Universität Jena, Germany
[3] Fraunhofer Institute for Applied Optics and Precision Engineering, Albert-Einstein-Str. 7, Jena, Germany
[4] Institute of Applied Physics and Abbe Center of Photonics, Friedrich-Schiller-Universität Jena, Germany
[5] ICREA, Pg. Lluís Companys 23, 08010 Barcelona, Spain
[*]Correspondence to: jens.biegert@icfo.eu
[†]These authors contributed equally

**Non-adiabatic dynamics and conical intersections play a central role in the chemistry of most polyatomic molecules, ranging from isomerization to heterocyclic ring opening and avoided photo-damage of DNA. Studying the underpinning correlated dynamics of electronic and nuclear wave packets is a major challenge in real-time and, many times involves optically dark transient states. We show that attosecond core-level spectroscopy reveals the pathway dynamics of neutral furan across its conical intersections and dark states. Our method measures electronic-nuclear correlations to detect the dephasing of electronic coherence due to nuclear motion and identifies the ring-opened isomer as the dominant product. These results demonstrate the efficacy of attosecond core level spectroscopy as a potent method to investigate the real-time dynamics of photochemical reaction pathways in complex molecular systems.**

Chemical reactions occur through the redistribution of valence electrons, which involves nuclear motion mediated by the Coulomb force. The coupling between electronic and nuclear degrees of freedom (*1*) can often be very strong, thus making the real-time investigation of electronic dynamics very challenging (*2*). However, such non-adiabatic dynamics are essential to understand as they frequently occur in polyatomic molecular systems. Conical intersections (*3*, *4*) provide fast and radiation-less energy relaxation between different potential energy surfaces, efficiently converting electronic excitation into vibrational excitation. Prominent examples are the retinal isomerization (*5*) of the vision process and the photo-stability of DNA bases (*6*). Due to the importance of valence electron dynamics, many methods were developed; e.g., ultrafast two-dimensional spectroscopy (*7–11*), attosecond transient absorption (*12–14*), time-resolved photoelectron spectroscopy (*15–17*), x-ray spectroscopy (*18*, *19*) and high harmonic spectroscopy (*20–25*). Nevertheless, a tremendous challenge for existing methods is still the combined requirement for ultrafast time resolution (*26–28*), the capability to disentangle time-overlapping electronic and nuclear dynamics (*29*) and to follow radiation-less decay across transient and optical dark states (*30*).

Here, we meet these combined challenges with attosecond core-level spectroscopy, which reveals the energy pathways and real-time conical intersection dynamics in photo-excited neutral furan and its vibrational dynamics on its native ultrafast time scale. The element and orbital selectivity of core-level spectroscopy, combined with attosecond temporal resolution, reveals the heterocyclic system's electronic and nuclear evolution predominantly into its ring-opening product. We chose to investigate furan ($C_4H_4O$) since it serves as a prototypical system for heterocyclic organic rings (*31, 32*), which are essential building blocks of polymers, fuels, pharmaceuticals, atmospheric constituents, organic electronics and light-harvesting systems, agrochemicals, and more. Thus, their photochemical dynamics and relaxation processes are of great interest (*33*) for various applications. The results of our investigation demonstrate that the initially recognized potential of ultrafast x-ray absorption spectroscopy (*34–39*) is now matured with attosecond core-level spectroscopy (*14, 34, 35, 40–42*) into a powerful analytical technique to elucidate complex chemical reaction dynamics.

Attosecond core-level spectroscopy provides information on unoccupied, more precisely, not-fully-occupied, electronic states via dipole-allowed core-level K or L shell transitions; thus, the method is sensitive to optically dark states that occur in many reaction pathways. Attosecond core-level spectroscopy has a vital distinction from general attosecond transient absorption spectroscopy (ATAS, (*12, 13, 43*)) in that the absorption spectrum arises only due to proper core transitions. I.e., 1s (K) or 2s, 2p (L) states avoid the final-state multiplet effects (*44*) of higher-laying states. The so ensured direct mapping between measured optical absorption to the electronic states of the system is critical for the unambiguous identification of electronic and nuclear effects (*45*).

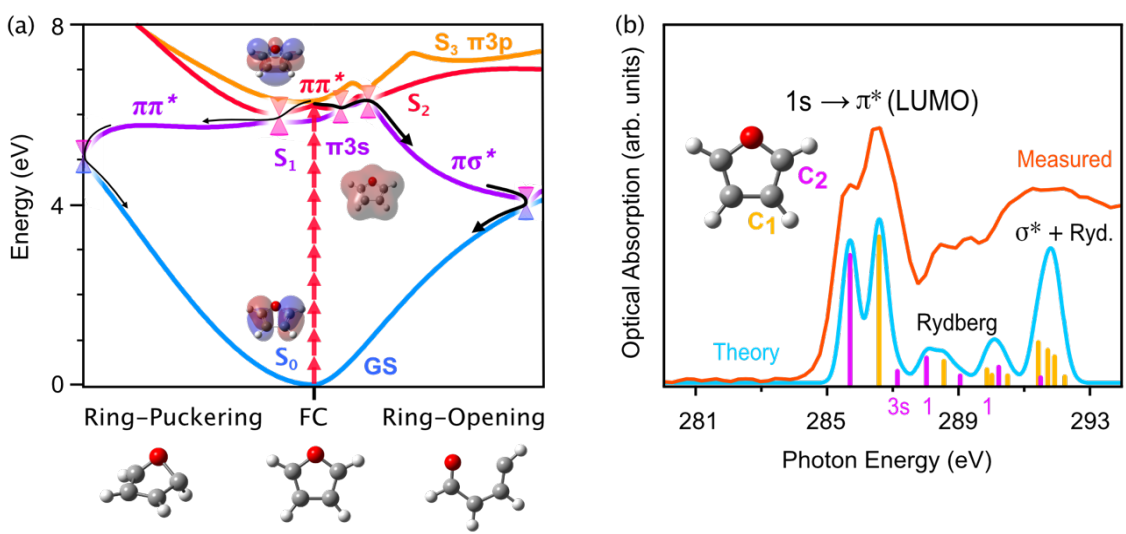

*Figure 1. (a) Potential energy surfaces showing the RO and RP pathways and the relevant electronic states. Note that the excited states S0, S1, S2, and S3 refer to the adiabatic states, which change their electronic configurations upon passage across the CIs. (b) Static (non-excited furan) carbon-edge XANES. The measured absorption (red) and the calculated one (blue) are shown. Purple and yellow vertical lines indicate spectral contributions from the two symmetry-distinguishable carbon atoms ($C_1$ and $C_2$). These contributions are due to the planar GS geometry with $C_{2v}$ symmetry (see inset of picture), leading to two distinct carbon atoms. Further indicated are contributions to Rydberg states and the anti-bonding $\sigma^*$ and higher order $\pi^*$ states.*

Relevant electronic states for photo-excited furan are shown in Fig. 1a. Infrared multi-photon absorption of 0.7-eV (1850 nm) photons excites the molecule from its electronic ground state (GS) to the $\pi\pi^*$ (S2) excited state. We note that the $\pi$3s (S1) Rydberg state is optically dark in the Franck-Condon (FC) region, while the $\pi$3p (S3) Rydberg state is energetically very close to the S2 state but possesses a 6x smaller oscillator strength.

Thus, predominantly the ππ* (S2) is excited. Theoretical investigations (*46*, *47*) have discussed two alternative pathways along which furan evolves after photo-excitation into the S2 state: (i) The molecule can evolve after ππ* excitation in the FC region via non-adiabatic passage through the S2(ππ*)/S1(ππ*) and S1(ππ*)/S0 conical intersections back to the initial GS (S0). This pathway leads to distortion of the nuclear framework out of the molecular plane, called ring-puckering (RP). (ii) Alternatively, ππ* excitation in the FC region leads to a transient population of the optically dark πσ* state via non-adiabatic passage through the S2(ππ*)/S1(πσ*) conical intersection. The anti-bonding πσ* state is repulsive, thus leading to the fission of the C-O bond. The system further relaxes through the S1(πσ*)/S0(πσ*) conical intersection to the electronic ground state. This non-adiabatic pathway is called ring-opening (RO). Both pathways (*48*) lead to strong vibrational excitation.

Previous experimental studies using multi-photon and single-photon excitation aimed to identify these pathways but found different end products. For instance, some photoemission experiments have identified RP (*49*, *50*) as the dominant pathway, while others identified RO as the primary pathway (*51*). Several theoretical investigations also favor RP as the main relaxation pathway (*52*). It is worth mentioning that recent investigations with ultrafast electron diffraction on cyclohexadiene (*53*) and x-ray absorption spectroscopy on furfural (*54*) contradict most photoemission studies by identifying RO as the main relaxation pathway for those heterocyclic ring systems. Thus far, despite recent progress (*42*, *54*), identifying the non-adiabatic pathways, the expected vibronic coherences, and the vibrational dynamics has been obscured by the challenge of temporally resolving and directly identifying them (*55*, *56*).

Here, we meet this challenge with IR-pump, core-level x-ray absorption near-edge spectroscopy (XANES) probe measurements with an attosecond soft x-ray beamline on gas-phase furan. We photo-excite furan inside a 4-mm-long effusive cell with 400-micron entrance and exit holes for pump and probe beams by multi-photon absorption of 17-fs-duration, carrier-envelope phase (CEP) stable pulses at 1850 nm (0.7 eV) with focused intensity up to 90 TW/cm$^2$. The attosecond soft-x-ray (SXR) pulse interrogates photo-excited furan in transmission by measuring the XANES with a homebuilt SXR spectrometer with 1/1000 resolution at the carbon K-edge at 284 eV; see the SI for experimental details.

Before detailing the temporal reaction dynamics, we highlight that the sensitivity of XANES to orbital symmetries and occupation already provides a direct way to distinguish between RO and RP. In the GS, the molecule exhibits C$_{2v}$ symmetry, i.e., only 2 of the 4 carbon atoms are distinct. This symmetry arises from the different binding of the two sets of carbon atoms (C1 and C2) and sensitively registers as two energetically-shifted core transitions C$_{1,2}$(1s)→π* in the x-ray absorption spectrum. Figure 1b shows the measured static (un-pumped) XANES, in which the two distinct carbon atoms are visible as double peaks around 285.6 eV and 286.5 eV. Further visible in the edge region, above the two LUMO peaks, are contributions from transitions into the Rydberg and σ* states and higher order and mixed π* states.

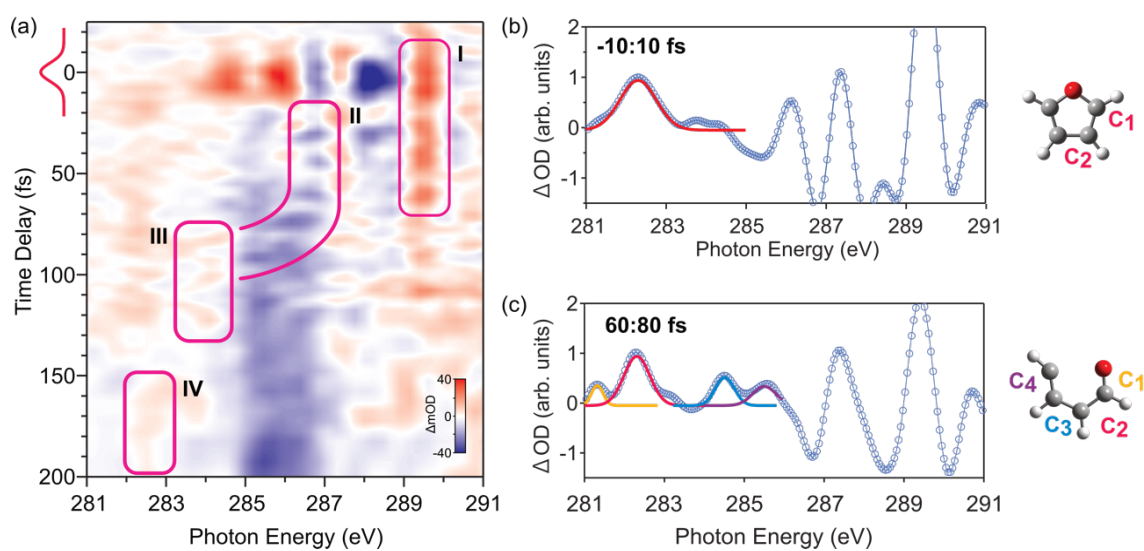

*Figure 2. (a) Differential attosecond-resolved XANES measurement. Indicated is the 17-fs pump pulse, whose envelope is centered at "0" delay time. Further shown are areas "I"-"IV," which facilitate discussion of different spectral contributions in the text. Substantial changes in absorption at zero delay time are observed, followed by various modulations to varying frequencies due to electronic and vibrational coherences. (b) Time-integrated lineouts across the energy range of (a) during the pump pulse and (c) after relaxation to the GS of RO. A singly occupied molecular orbital (SOMO) peak (b) in the pre-edge region, below 285 eV, indicates excitation of the system into the LUMO. Carbon-heteroatom bond fission (RO) breaks the $C_{2v}$ symmetry and, consequently, the excitation across 4 distinct carbon atoms ($C_1$-$C_4$) registers as splitting of the singly-occupied molecular orbital (SOMO) peak into 4 absorption peaks (c). See text and SI for details.*

We now turn to investigate the reaction dynamics of photo-excited furan and will follow its time evolution. Figure 2a shows the measured differential absorption spectrum, which displays the changes in the XANES relative to un-excited (ground-state) furan. In the following, we will use the combination of experiment and theory to elucidate the entire non-adiabatic evolution of the furan system. Our theoretical methods include coupled cluster singles and doubles calculations (fc-CVS-EOM-EE-CCSD), time-evolution with ab-initio on-the-fly-surface hopping, and non-adiabatic and laser couplings; see the SI for detailed information.

### 1st conical intersection S2(ππ*)/S1(π3s)

Immediately apparent in Fig. 2a at early delay times are strong modulations in the differential absorption spectrum, both positive (red) and negative (blue), and across the entire shown photon energy range. Optical excitation of one electron from the highest occupied molecular orbital (HOMO) into the lowest unoccupied molecular orbital (LUMO) creates singly-occupied molecular orbitals (SOMO). It registers as a new absorption peak in the differential absorption spectrum. Thus, the appearance of the single SOMO peak in the pre-edge region (below 285 eV), shown as a line out in Fig. 2b, is due to electronic excitation of the ππ* (S2) state in the FC region and core excitation of $C_{1,2}$(1s) into the hole created in the π orbital. From a comparison with the static XANES (Fig. 1b), we find that the additional modulations in the edge region, above 285 eV, are due to the simultaneous excitation to the ππ* (S2) and π3p (S3) states, and their coupling with the π3s (S1) dark state. Consequentially, an analysis of the time evolution of different energetic signatures above the edge region provides insight into the electronic coupling and dynamics between those states: Figure 3a shows the time evolution at 287 and 289.5 eV, covering positive and negative signal regions „I" and "II" in Fig. 2a. The two differential absorption curves directly reveal a rapid exchange of electronic population, which manifest as the rise of either signal that maximizes in about 30 fs. We note that the timed delay axis was chosen such that time "0" coincides with the maximum of the pump pulse envelope. Already within the 17-fs-FWHM pump pulse, strong modulations are visible. A Fourier analysis reveals a modulation frequency of 63 $\pm$ 9 THz, corresponding to a period of 16 fs; see inset in Fig. 3a. Interestingly, these modulations are in anti-phase until about 80 fs, after which their phase difference changes towards in-phase motion.

To elucidate the origin of this behavior, Fig. 4a shows the calculated population dynamics and the evolution of electronic states. The remarkable agreement with the measurement allows us to explain the observed behavior: We find from the theory that already after 12 fs, ππ* population traverses the S2(ππ*)/S1(π3s) conical intersection. Passage through the conical intersection maintains the states' electronic character and establishes a fixed phase relation for the evolution of electronic excitation along the two different potential energy surfaces (PES) ππ* (S2) and π3s (S1), which manifests in quantum beats of the electronic coherence. The measured beat frequency of 63 $\pm$ 9 THz is in excellent agreement with theory, which predicts frequencies of 60 $\pm$ 6 THz for RO and 62 $\pm$ 9 THz for RP; see Fig. 4c.

Further, the coherent motion of charge density across the two distinct carbon atoms of furan exhibits a π-phase shift between the spatially separate nuclear sites, shown as red and green insets in Fig. 3a. We find that this phase shift registers via the complex transition dipole matrix element of the core-transition to the common final (valence) state as anti-phase evolution of quantum beats at the two energies that are distinct for the two carbon sites. The detection of the beating signifies that the measurement provides site-specific tracking of the coherent wave-packet evolution across the molecule's nuclear structure and clocking electronic coherence between different states.

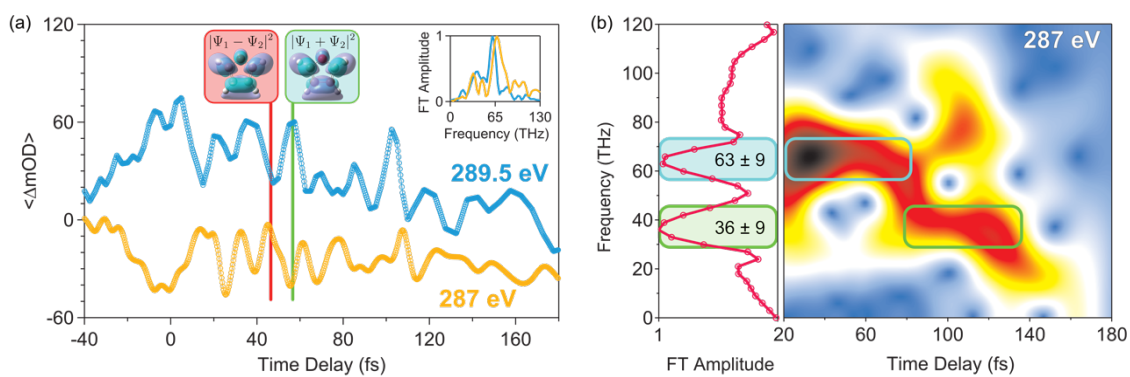

*Figure 3. (a) Lineout data from the differential XANES (Fig. 2a) at two distinct energies for negative and positive differences. After the buildup of the negative (yellow) and positive (blue) absorption signals, electronic quantum beating at 64 THz is observed. After the initial rise during the 17-fs-duration pump*

*pulse, electronic coherences beat in anti-phase due to coherent wave-packet evolution across the molecule's nuclear structure with two distinct carbon sites. The relevant wave packet superpositions are shown inside the red and green panels above the curves. After 80 fs, the phase relation changes to in-phase beating, which decays after 140 fs. (b) A windowed Fourier analysis of the 287-eV curve in (a) reveals the time evolution of frequencies; analysis of the other curve yields comparable data. The persistence of a frequency over a time range reveals coherence. In the case of furan, we find that the electronic coherence, with 63 ± 9 THz beat frequency, dephases due to the buildup of vibrational excitation at a lower frequency of 36 ± 9 THz.*

### 2$^{nd}$ conical intersection S2(ππ*)/S1(πσ*) and ππ*

Following the dynamic evolution of the electronic coherence at 63 THz, we observe its dephasing towards in-phase motion at 80 fs and its overall decay after 140 fs. Intriguingly, the dephasing is accompanied by the appearance of a low-frequency mode at 36 ± 9 THz (Fig. 3b), which persists until 140 fs. The theory yields that the ππ* excitation passes through another conical intersection, S2(ππ*)/S1(πσ*), after 58 fs making the S1 state the most strongly populated electronic state (Fig. 4a). The appearance of the 36 ± 9 THz frequency mode originates from nuclear motion. Performing a normal mode analysis on the ensemble of trajectories, we indeed identify a 37 THz mode as being unique to the RO pathway (see Fig. 4d). This finding is explained by the stretching of the nuclear framework, which destabilizes the S2 and S3 PES. At the same time, it stabilizes S1 (see Fig. 4b and c). Moreover, our combined analysis yields that non-adiabatic passage through the S2(ππ*)/S1(πσ*) conical intersection results in a change of the electronic character to the optically dark dissociative πσ* state after 58 fs. This change is accompanied by the dephasing of the coherent electronic wave-packet motion due to the rapidly increasing energy separation of electronic states on the RO trajectory (Fig. 4b). In fact, the increasing PES gradient of the RO trajectory results in strong deformation of the nuclear framework and coherent vibrational excitation. This is visible as a large chemical shift from 286.5 to 284 eV in the differential absorption spectrum; see Fig. 2a area „II"-"III". The accompanying observed splitting of the single SOMO peak (Fig. 2c) into 4 peaks after about 70 fs is the unmistakable signature of ring scission due to the breakage of $C_{2v}$ molecular symmetry.

### 3$^{rd}$ conical intersection S1(πσ*)/S0

Lastly, we find that the SOMO peaks decay after about 140 fs. This is in excellent agreement with the theory which predicts passage along the RO trajectory through the S1(πσ*)/S0(πσ*) conical intersection (Fig. 2a, area "IV") after 158 fs. The combined observation of the SOMO peak splitting (Fig. 2c), the time scales of electronic coherences (Fig. 3a), and the measured vibrational coherence with a frequency of 36 ± 9 THz (c.f. Fig. 3b and Fig. 4d) are clear evidence of the system's evolution towards RO. Theory predicts a probability of 76% for the observed relaxation channel along the RO coordinate. This is in excellent agreement with the measurement.

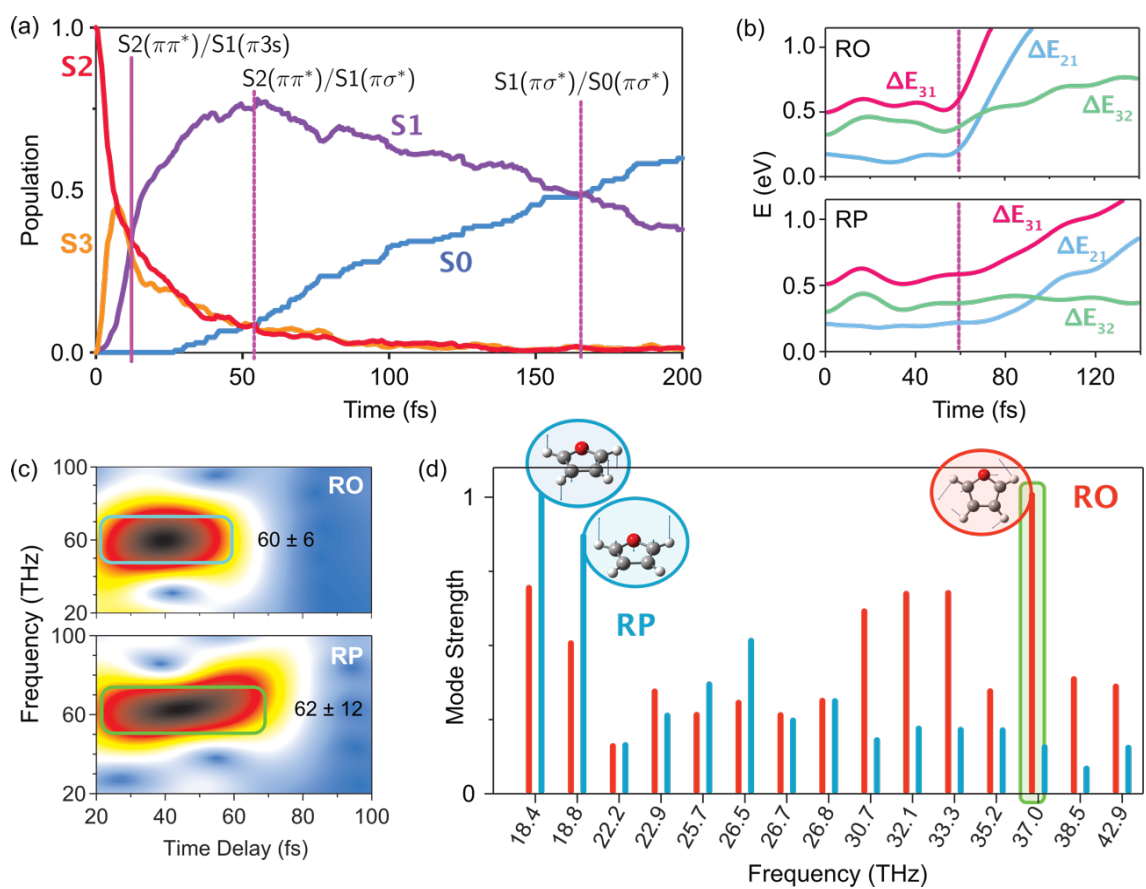

*Figure 4. (a) Theoretical results showing the time evolution of the relevant three excited states S1-S3 together with the S0 ground state. The vertical dashed lines (purple) indicate the temporal locations of non-adiabatic couplings between the various PES: S2($\pi\pi^*$)/S1($\pi3s$) CI at 12 fs, S2($\pi\pi^*$)/S1($\pi\sigma^*$) CI at 58 fs and S1($\pi\sigma^*$)/S0 CI at 158 fs. (b) shows energy differences between excited electronic states for the RO (top) and RP (bottom) pathways. In accord with the measurement, we find a substantial divergence of $\Delta E_{31}$ and $\Delta E_{21}$ for the RO pathways explains the dephasing of the electronic coherence after about 60 fs. (c) shows a windowed Fourier analysis of calculated electronic coherences. The analysis reveals indeed electronic coherence, which persists about 60 fs with a 60 THz mode for RO and 62 THz modes for RP, respectively. A normal mode analysis (d) in comparison with measurement settles that the RO pathway dominates the reaction dynamics. The 37.0 THz RO mode from theory is in excellent agreement with the measurement of $36 \pm 9$ THz (Fig. 3b). Further indicated in the blue and red panels next to the modes are the relevant RP and RO modes.*

We have shown that attosecond core-level spectroscopy is a potent method to reveal correlated multi-body dynamics in a complex molecular system on its native ultrafast timescale. We demonstrate the technique provides several different observables that allow us to disentangle the entire time-evolution of complex-coupled molecular electronic and vibrational dynamics after photo-excitation. The method's particular sensitivity to electronic coherences and phases and optically dark states and vibrational dynamics allows to resolve and identify non-adiabatic passages, changes in a system's electronic character, and electronic state switching. Further, core-level spectroscopy is state and element-specific, permitting spatially resolving electronic wave-packet dynamics across the molecular framework. This study on furan disentangles the system's whole non-adiabatic passage across three conical intersections and reveals the entire buildup of electronic coherences with their dephasing and vibrational cooling to the ring-opening ground state. Such combined energy and time resolution of attosecond core-level spectroscopy reveals hitherto inaccessible insight into molecular systems' real-time electronic and nuclear dynamics. Armed with intricate knowledge of a molecular system's different couplings and phase evolution, this may provide a decisive new basis for an engineered approach to quantum control. Insight into vibronic couplings may allow arresting dephasing of electronic excitations to facilitate efficient molecular reaction dynamics, to study the correlated multi-body dynamics leading to isomerization (Azo and retinal), or to understand energy relaxation dynamics in DNA.

Acknowledgment

J.B. acknowledges financial support from the European Research Council for ERC Advanced Grant "TRANSFORMER" (788218), ERC Proof of Concept Grant "miniX" (840010), FET-OPEN "PETACom" (829153), FET-OPEN "OPTOlogic" (899794), FET-OPEN "TwistedNano" (101046424), Laserlab-Europe (871124), Marie Skłodowska-Curie ITN "smart-X" (860553), MINECO for Plan Nacional PID2020–112664 GB-I00; AGAUR for 2017 SGR 1639, MINECO for "Severo Ochoa" (CEX2019-000910-S), Fundació Cellex Barcelona, the CERCA Programme/Generalitat de Catalunya, and the Alexander von Humboldt Foundation for the Friedrich Wilhelm Bessel Prize. S.S. acknowledges Marie Skłodowska-Curie Grant Agreement No. 713729 (COFUND). M.R. and A.S. acknowledge Marie Skłodowska-Curie Grant Agreement No. 754510 (PROBIST). K.M.Z. and S.G. are part of the Max Planck School of Photonics supported by BMBF, Max Planck Society, and Fraunhofer Society. S.G. highly acknowledges support from the European Research Council via the Consolidator Grant QUEM-CHEM (772676) and the CRC 1375 NOA – "Nonlinear Optics down to Atomic scales". We thank J. Menino and C. Dengra for their technical support.